# Cache Replacement Policies for Cooperative Caching in Mobile Ad hoc Networks

Preetha Theresa Joy[1] and K. Poulose Jacob[2]
[1,2] Dept. of Computer Science, Cochin University of Science and Technology,
Kochi, Kerala, India.

**Abstract**

Cooperative caching is a technique used in mobile ad hoc networks to improve the efficiency of information access by reducing the access latency and bandwidth usage. Cache replacement policy plays a significant role in response time reduction by selecting suitable subset of items for eviction from the cache. In this paper we have made a review of the existing cache replacement algorithms proposed for cooperative caching in ad hoc networks. We made an attempt to classify existing replacement policies for ad hoc networks based on the replacement decision taken. In addition, this paper suggests some alternative techniques for cache replacement. Finally, the paper concludes with a discussion on future research directions.

*Keywords:* Data Caching, Cache Replacement, MANET, Cooperative caching.

## 1. Introduction

Wireless mobile communication is a fastest growing segment in communication industry. It has currently supplemented or replaced the existing wired networks in many places. The wide range of applications and new technologies simulated this enormous growth. The new wireless traffic will support heterogeneous traffic, consisting of voice, video and data. Wireless networking environments can be classified in to two different types of architectures, infrastructure based and ad hoc based. The former type is most commonly deployed one, as it is used in wireless LANS and global wireless networks. An infrastructure based wireless network uses fixed network access points with which mobile terminals interact for communication and this requires the mobile terminal to be in the communication range of a base station. The ad hoc based network structure alleviates this problem by enabling mobile terminals to cooperatively form a dynamic network without any pre existing infrastructure. It is much convenient for accessing information available in local area and possibly reaching a WLAN base station, which comes at no cost for users.

Mobile terminals available today have powerful hard ware, but the capacity of the batteries goes up slowly and all these powerful components reduce battery life. Therefore adequate measures should be taken to save energy. Communication is one of the major sources of energy consumption. By reducing the data traffic energy can be conserved for longer time. Data caching has been introduced as a techniques to reduce the data traffic and access latency. By caching data the data request can be served from the mobile clients without sending it to the data source each time. It is a major technique used in the web to reduce the access latency. In web, caching is implemented at various points in the network. At the top level web server uses caching, and then comes the proxy server cache and finally client uses a cache in the browser.

This paper provides a general comparison of the cache replacement policies used in mobile ad hoc networks based on the criteria used for evicting documents. We reviewed the various replacement policies used in cooperative caching in ad hoc networks. The topic of caching in ad hoc networks is rather new, and not much work has been done in this area. We classified the replacement policies for MANETs in to two groups uncoordinated and coordinated. In uncoordinated replacement, the data item to be evicted is determined independently by each node based on its local access information. In coordinated replacement policy the mobile nodes which forms cooperative cache collectively takes the replacement decision.

## 2. Cache Replacement

Caching in wireless environment has unique constraints like scarce bandwidth, limited power supply, high mobility and limited cache space. Due to the space limitation, the mobile nodes can store only a subset of the frequently accessed data. The availability of the data in local cache can significantly improve the performance since it overcomes the constraints in wireless environment. A good replacement mechanism is needed to distinguish the items to be kept in cache and that is to be removed when the cache is full. While it would be possible to pick a random object to replace when cache is full, system performance will be better if we choose an object that is not heavily used. If a heavily used data item is removed it will probably have to be brought back quickly, resulting in extra overhead. So a good replacement policy is essential to achieve high hit rates.

The extensive research on caching for wired networks can be adapted for the wireless environment with modifications to account for mobile terminal limitations and the dynamics of the wireless channel.

## 2.1 Performance Metrics

Caching in wireless networks deals with data items of different costs and sizes. Performance measures used should consider this non uniformity. It is not possible to identify a best replacement policy as different schemes uses different optimization strategy.

The typical measures used to analyze the cache replacement policy are hit ratio, byte hit ratio and delay savings ratio [1]. Let $S_i$ be the size of the data item i, $C_i$ the cost of fetching the data item i into the cache, $R_i$ the total number of references made to data item i, $H_i$ the number of hit references made to data item i and $D_i$ is the delay time to fetch the data item i from the original data source to cache. Cache hit ratio defines the number of references made from the cache over the total number of references. It is a metric used in traditional caching systems like operating systems and database which handles data of uniform size, which may not be reliable metric for data items with varying size and cost. Byte hit ratio represents the number of bytes saved from retransmission by using the cache over the total number of bytes referenced. Delay savings Ratio represents the reduced latency by using a cache over the total latency when cache is not used. Due to the inconsistency in download time due to traffic variations, performance results based on this metric may vary.

$$HR = \sum H_i / \sum R_i$$
$$BHR = \sum S_i . H_i / \sum S_i . R_i$$
$$DSR = \sum D_i H_i / \sum D_i . R_i$$

## 3. Cache Replacement Policies in Wireless Networks

The cache replacement policies for wireless networks can be broadly classified in to function based, location based and semantic based policies based on the parameters considered for evicting data when cache is full.

## 3.1 Function Based Cache Replacement

The different parameters which affect the cache replacement for wireless mobile environments are data access pattern, access costs, mobility pattern, connectivity, bandwidth, update rates, location dependence of the data. In function based cache replacement each data item is assigned a value using a value function by combining these parameters. The data item or items with least value are chosen for replacement. Traditional cache replacement policies like LRU and LFU use only one parameter to make the replacement decision.

## 3.2 Location Based Cache Replacement

In Location Dependent information services (LDIS) the value of the data item depends on the location and varies as the user changes his location. The factors considered in a location aware replacement policy are the valid scope area, distance and direction of client movement. The area under which the data item is valid is the valid scope area. Distance is the distance between mobile node's current location and the valid scope area. When the data is distant from the valid scope area, it will have a lower chance to become useful. Direction indicates the direction of data movement from the valid scope area. The data that are moving in the opposite direction of the valid scope area will be irrelevant after sometime.

## 3.3 Semantic Cache Replacement

In semantic caching each mobile client have a semantic description of data and query processing make use of this data. Semantic caching is ideal for location dependent services as the probability of having similar queries is high. There may be a large degree of overlap in the results of consecutive queries. If the results are kept in a cache, part of the new query result could be answered from the local cache. The data needed from the server can be retrieved by specifying a reduced query thereby reducing the network traffic. Very few replacement strategies are proposed for semantic caching.

## 4. Cache Replacement Policies in Ad hoc networks

Data caching in MANET is proposed as cooperative caching. In cooperative caching the local cache in each node is shared among the adjacent nodes and they form a large unified cache. So in a cooperative caching environment, the mobile hosts can obtain data items not only from local cache but also from the cache of their neighboring nodes. This aims at maximizing the amount of data that can be served from the cache so that the server delays can reduced which in turn decreases the response time for the client. In many applications of MANET like automated highways and factories, smart homes and appliances, smart class rooms, mobile nodes share common interest. So sharing cache contents between mobile nodes offers significant benefits.

Cache replacement algorithm greatly improves the effectiveness of the cache by selecting suitable subset of data for caching. The available cache replacement mechanisms for ad hoc network can be categorized in to coordinated and uncoordinated depending on how replacement decision is made. In uncoordinated scheme the replacement decision is made by individual nodes. In order to cache the incoming data when the cache is full, replacement algorithm chooses the data items to be removed by making use of the local parameters in each node.

### 4.1 Uncoordinated Cache Replacement

In uncoordinated replacement strategy, in order to make the replacement decision when the cache is full, only the local parameters are considered. In cooperative caching this may not be desirable. In the following section we discuss various uncoordinated cache replacement policies for mobile ad hoc networks.

LRU

LRU (Least Recently Used) replacement policy is based on the observation that data that have been frequently used recently will probably be used again in the future. Conversely, data that have not been used for ages will probably remain unused for a long time. In LRU when cache is full the data item that has been unused for the longest time is thrown out. It is a widely used algorithm in cache replacement. Logically, the cache consists of a list with most recently referenced data being in the front of the list. Upon each reference the data is moved from its existing position to the front of the list. When a new data comes in it is placed on the top of the list and the data at the back end is removed. LRU doesn't take in to account the non uniformity in the size of data, which is an important factor in mobile communication as the cost to fetch the data depends on size.

LRU Min

LRU Min [2] is a variant of LRU that tries to minimize the number of documents replaced. It is similar to LRU in implementation but will consider size of the data during replacement. In this scheme the data is arranged on the basis of access time and if a data item of size S needs to be cached it will search for items least recently accessed with size greater than S. If there isn't any data in cache with size S, we start removing the items with size greater than S/2 and then objects of size S/4 until enough cache space is created. LRU Min policy will increase the hit ratio of smaller sized data items.

SXO

This is a local replacement policy [3] which considers the parameters data size and access frequency for replacement. Here larger sized data items are removed first as they occupy more cache space. More cache space can be made available by replacing bigger objects. The second parameter considered is order (di) which gives the frequency of access of data. Here replacement is done by combining the two parameters as value $(di)=S*order(di)$. The advantage of this scheme is that the parameters used are easily available. But recently accessed data are not given any privilege.

LUV

A cache replacement policy based on least utility value (LUV) has been used in [4] For computing the LUV of a data item the access probability ($A_i$), size of the data item ($S_i$), coherency which can be known by $TTL_i$ field and distance ($\delta$) between the mobile client and data source were considered. Eq. for $utility_i$ function for a data item (di) is:

$utility_i = A_i \cdot TTL_i \, \delta_i / S_i$

### 4.2 Coordinated Cache replacement Policies

Coordinated cache replacement strategy for cooperative caching schemes in mobile environments should ideally consider cache admission control policy. Cache admission control decides whether the incoming data is cacheable or not. Substantial amount of cache space can be saved by proper admission control, which can be utilized to store more appropriate data, thereby reducing the number of evictions. If a node doesn't cache the data that adjacent nodes have it can cache more distinct data items which increase the data availability. There is coordination between the neighboring nodes for the proper placing of data. Another feature of coordinated replacement is that the evicted data may be stored in neighboring nodes which have free space. Some of the replacement policies which make use of coordinated cache replacement are given below.

TDS

The cache replacement [5] is based on two parameters distance (D) which is measured as the number of hops and access frequency. As the network is mobile the value of distance (D) may become obsolete. So the value is chosen based on the time at which it is last updated. The T value is obtained by the formula $1/t_{cur} - t_{update}$. Distance is updated by looking at the value of T. Based on how the distance and time is selected three different schemes are proposed TDS_D, TDS_T and TDS_N. TDS_D considers distance as the replacement criteria. If two data items have the same distance least value of (D+T) is replaced. In TDS_T the replacement decision is made by selecting the data with lowest T value. In the third scheme product of distance and access frequency is considered. In these algorithms TDS_D has the lower success rate and TDS_T has the higher hit ratio.

LUV Mi

This replacement scheme [6], has two parts replacement and migration. The replacement decision is based on a utility value formed by combining the parameters access probability, distance, size and coherency. In the migration part the replaced data is stored in the neighboring nodes which have sufficient space. For migration the data with highest utility value is given preference. Here even though the replacement decision is made locally migration is a coordinated operation. In order to save the cache space the data item is cached based on the location of the data source. If it is from the same cluster the data is not cached. The limitation of this scheme is that no checking is done whether the data is already present in the migrating node.

ECORP

Energy efficient cooperative cache replacement problem (ECORP) [7] is an energy efficient cache replacement policy used in ad hoc networks. Here the replacement decision is done based on the energy utilization for each data access. For this, they considered the energy for in zone communication, energy for sending the object, energy for receiving and energy cost for forwarding the object. Based on this they proposed a dynamic ECORP DP and ECOPR _greedy algorithms to replace data. The neighboring nodes will not cache the same data item in its local cache which reduces the redundancy and increases hit ratio.

Count Vector
In this scheme [8], each data item maintains a count which gives the number of nodes having the same data. Whenever the cache is full, data item with maximum count is removed first as this will be available in the neighboring nodes. Whenever a data item is removed from the cache the access count will be decremented by one. Initially when the data is brought in to cache the count is set to zero. Table1 shows a comparison of different cache replacement policies.

## 5. Discussion and Future work

Most of the replacement algorithms used in ad hoc networks is LRU based which uses the property of temporal locality. This is favorable for MANET which is formed for a short period of time with small memory capacity. Frequency based algorithms will be beneficial for long term accesses. It is better if the function based policies can adapt to different workload condition. In these schemes if we are using too many parameters for finding the value function, which are not easily available the performance can be degraded. Most of the replacement algorithms mentioned above uses cache hit ratio as the performance metric. In wireless network the cost to download data item from the server may vary. So in some cases this may not be the best performance metric. Schemes which improve cache hit ratio and reduce access latency should be devised. In cooperative caching coordinated cache replacement is more effective than local replacement since the replacement decision is made by considering the information available in the neighboring nodes. The area of cache replacement in cooperative caching has not received much attention. Lot of work needs to be done in this area to find better replacement policies. Location dependent services are becoming popular in ad hoc networks. Replacement policies which consider location dependent parameters should be devised for cooperative caching in ad hoc networks. Another area of research in ad hoc networks is semantic caching in which the query is served from the cache based on the semantic description and results of previous queries. Cache admission control also plays role in improving the performance of cooperative cache. Value based admission control can be incorporated to minimize the number of replacements. Cache replacement based on Quality of Service (QOS) parameters can be explored. An alternative to cache replacement is that the data items that have their Time to Live (TTL) expired can be removed as the data becomes stale and cannot be used. So periodical checks can be done to delete the data items with TTL expired.

Table.1 Comparison of cache replacement policies for cooperative caching

| Algorithm | Parameters Considered | Eviction | Performance measure | Advantage | Disadvantage |
|---|---|---|---|---|---|
| LRU | Last access time | Least access time | Cache hit Ratio | Simple to implement temporal based | Non uniformity in data is not considered |
| LRU Min | Last Access time, Size | Lowest access time, size greater than incoming data | Average query delay | Cache hit ratio of smaller sized objects is increased. | Not suitable for large caches. |
| SXO | Size and access frequency | Largest size, low access frequency | Cache hit ratio | Parameters used are easily available. | Recency of data is not considered. |
| LUV | Access probability, size, coherence and distance. | Low access probability, bigger size, low consistency, lowest distance | Byte hit ratio, average query latency, message overhead | Considers all the temporal data. | Distance parameter may vary and is not considered. |
| TDS | Distance and access frequency | Low access rate and lowest distance | Success rate, cache hit ratio. | Value of distance is updated. | Doesn't consider recency of data item. |
| LUV Mi | Access probability, size, coherence and distance. | Low access probability, bigger size, low consistency, lowest distance | Byte hit ratio, average query latency, message overhead | Evicted data is stored in adjacent nodes. | No checking is done before storing data. |
| ECORP | Energy for in zone communication, sending object, receiving object | Lowest energy value | Cache hit ratio, average access delay | Energy is taken as the important parameter | Computing energy for each task is not easy |
| Count Vector | Access count | Maximum access count | Average access time | Coordinated simple to implement | Data redundancy is high. |

## 6. Conclusion

In this paper we made a general comparison of the major replacement policies in wireless networks and summarized the main points. Numerous replacement policies are proposed for wireless networks, but a few for cooperative caching in ad hoc networks. We also summarized the operation, strengths and drawbacks of these algorithms. Finally we provided some alternatives for cache replacement and identified topics for future research.